\documentclass[aps,prl,reprint,numerical,superscriptaddress]{revtex4-1}
\usepackage{pslatex,graphicx,verbatim, amsmath,amssymb,booktabs}
\usepackage[caption=false]{subfig}
\usepackage[inline]{enumitem}
\usepackage[breaklinks=true,
colorlinks = true,
citecolor=blue, 
linkcolor=blue,
urlcolor=blue]{hyperref}
\usepackage{csquotes}
\newcommand{\vect}[1]{\textbf{#1}}
\usepackage[T1]{fontenc}
\usepackage{ae,aecompl}
\usepackage{times}
\begin{document}
\title{ Using random numbers to obtain Kohn-Sham potential for a given density}
\author{Ashish Kumar}
\email[]{ashishkr@iitk.ac.in}
\affiliation{Department of Physics, Indian Institute of Technology Kanpur, Kanpur-208016, India}       
\author{Manoj K. Harbola}
\email[]{mkh@iitk.ac.in}
\affiliation{Department of Physics, Indian Institute of Technology Kanpur, Kanpur-208016, India}
\date{\today}      	
\begin{abstract}
Most  of the density-to-potential inversion methods developed over the years follow a general algorithm
 $v_{xc}^{i+1}(\vect{r}) = v_{xc}^{i}(\vect{r}) + \Delta v_{xc}(\vect{r})$, where $\Delta v_{xc}(\vect{r}) = \frac{\delta S[\rho]}{\delta \rho(\vect{r})}\Big|_{ \rho_i(\vect{r})} - \frac{\delta S[\rho]}{\delta \rho(\vect{r})}\Big|_{ \rho_0(\vect{r})}$ and $S[\rho]$ is an appropriately chosen density functional. In this work we show that this algorithm can be used with random numbers to obtain the exchange-correlation potential for a given density. This obviates the need to evaluate the functional $S[\rho]$ in each iterative step. The method is demonstrated by calculating  exchange-correlation potential of atoms, clusters and the Hookium.
\end{abstract}
\maketitle
\par Density functional theory (DFT) \cite{Hohn,Kohn_1965,Yang,Gross,Drei} is the most widely used theory of electronic structure \cite{Burke_1_An_rev_2015}. Although exact in principle, its implementation requires making approximations, which have been become better and better with the time \cite{Perdew_86,Becke_1988,Lee_PRB.37.785,Perdew_PRL.77.3865,Perdew_PRL.82.2544, Sun_PNAS,Sun_PRL.115.036402,SCANACC}. Developing accurate exchange-correlation functionals has therefore been and continues to be an active area of research in DFT \cite{Jianmin_PRL.91.146401} . On the other hand, knowing the exact results, wherever possible,  for  an operationally approximate theory is also of paramount importance. These results can be used to put the approximations made in proper perspective. Consequentially it this can lead to improving the approximations employed.
\par Most of the DFT calculations are performed using its Kohn-Sham (KS) formulation \cite{Kohn_1965}. In this method the density $\rho(\vect{r})$ of an  \text{N-electron} system is expressed in terms of independent particle orbitals $\{\phi_i \}$ as
\begin{equation}\label{eq3.1.1}
\rho(\vect{r}) = \sum f_i| \phi_i(\vect{r})|^2,
\end{equation}
where $\{f_i\}$ are the occupation numbers of these orbitals in the ground-state configuration. The orbitals are obtained by solving the KS equation (atomic units are used throughout)
\begin{equation}\label{eq3.1.2}
\left[ -\frac{1}{2} \nabla^2 + v_{ext}(\vect{r}) +v_H(\vect{r}) +v_{xc}(\vect{r})\right]\phi_i(\vect{r}) = \epsilon_i\phi_i(\vect{r})
\end{equation}
self-consistently. Here $v_{ext}(\vect{r})$ is the external potential in which electrons are moving,
\begin{equation}\label{eq3.1.3}
v_H(\vect{r}) = \int \frac{\rho(\vect{r}')}{|\vect{r}-\vect{r}'|}d\vect{r}'
\end{equation}
is the Hartree potential and $v_{xc}(\vect{r})$ is the exchange-correlation potential. In developing KS theory, both Hartree and the exchange-correlation potential are obtained as functional derivatives of the corresponding energy functionals viz. the Hartree energy functional
\begin{equation}\label{eq3.1.4}
E_H[\rho] = \frac{1}{2}\int \frac{\rho(\vect{r})\rho(\vect{r}')}{|\vect{r}-\vect{r}'|}d\vect{r}'d\vect{r}'
\end{equation}
and the exchange-correlation energy functional $E_{xc}[\rho]$. As is well understood,  the exchange-correlation energy functional $E_{xc}[\rho]$ is not known exactly and has to be approximated. Thus in implementing  KS-DFT, both the exchange-correlation energy functional and the exchange-correlation potential are treated approximately. 
\par In developing  exchange-correlation functionals, exactly known results about these quantities help in making them better. For example, development of initial GGA exchange functional by Perdew \cite{Perdew_86} made use of the exact sum rule satisfied by the exchange hole while that by Becke \cite{Becke_1988} employed the asymptotic behaviour of the exact exchange energy density. In the context of present work, two exact properties we mention are the exact asymptotic behaviour of the exchange-correlation potential and the ionization-potential theorem. For finite systems $v_{xc}(\vect{r})$ goes  as $-\frac{1}{r}$  as a function of distance $r$ from the system and for metallic surfaces it behaves as the image potential \cite{UvBarth_1985}. The second example is that of ionization potential theorem which states that the value of the highest occupied orbital energy $\epsilon_{max}$ of a system is equal to the negative of its exact ionization potential \cite{PPLB, LPS_1984}.
\par As is clear from the discussion above, the only component of KS potential that is not known  exactly is the exchange-correlation potential. This has led to various investigations into  providing the exact exchange-correlation potential and understanding its nature \cite{Buijse_1989,Gritsenko_1996,Teal2,Teal3,Teal4,Makmal_2011,Wagner_2012,2014_Gould,Kohut_2016,Proetto_2016,Godby_PRA_2016, Rabeet_2017,Staroverov_PNAS_18,kummel_18,2019_Gould}. Thus method for getting the exact exchange-correlation potential from the many-body wavefunction  \cite{Gritsenko_1998,Schipper_1998,Viktor_2013, Viktor_2015} or by inverting the exact density \cite{Werden, Stott_1988,Gorling_1992, Zhao_1992,Wang_1993, Zhao_1993, Zhao_1994, Wang_1993, Vlb_1994, Schipper1997, WY2,Peirs_2003, Stott_2004, Wagner_2014,Hollins_2017,Wasserman_2017,Finzel2018}  wherever these are available have also been developed over the years. This has led to insights into the behaviour of the exact exchange-correlation potential. Thus development of methods for getting the exact $v_{xc}(\vect{r})$ from a given density is an important research activity. In our recent work \cite{Kumar_2019}, we have demonstrated that the unified nature of various methods \cite{Werden, Stott_1988,Gorling_1992, Zhao_1992,Wang_1993, Zhao_1993, Zhao_1994, Wang_1993, Vlb_1994, Schipper1997, WY2,Peirs_2003, Stott_2004, Wagner_2014,Hollins_2017,Wasserman_2017,Finzel2018}  proposed  and have given a general algorithm that accomplishes this. Based on the insights provided in that work, in this paper we develop a purely numerical method for getting the exchange-correlation potential from a given density by using random numbers. In doing so, we make use of  Lieb's  definition  \cite{Lieb_1983} for the universal functional of DFT and maximize the related functional using random numbers. In the following we begin  by  briefly reviewing the  method of using Lieb's definition to invert a given density for getting the corresponding exchange-correlation potential. We then discuss its general nature \cite{Kumar_2019}. This generalization makes it possible to use random numbers to generate the exchange-correlation potential for a given density, thereby providing a novel approach to get the exchange-correlation potential. The corresponding results are then presented for a few systems. 
\begin{figure*}
\begin{center}
\includegraphics[width=.32\linewidth]{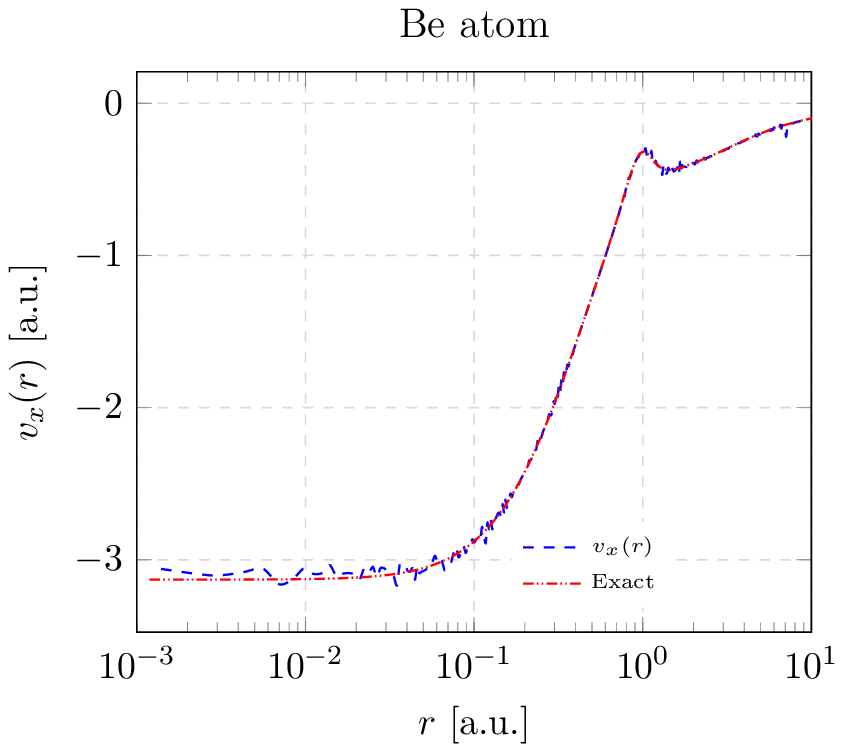}\hfil
\includegraphics[width=.32\linewidth]{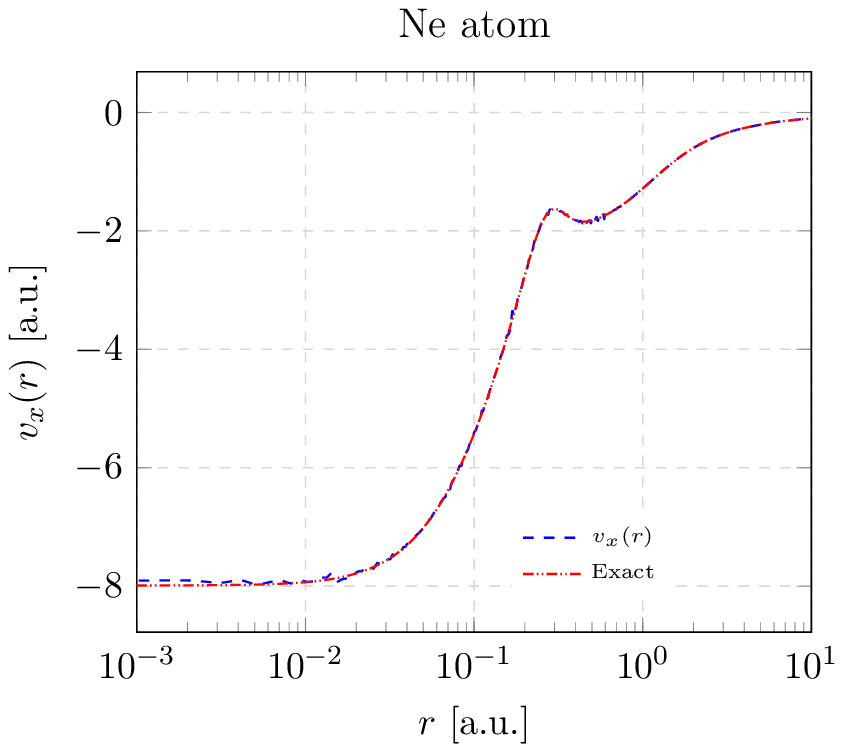}\hfil
\includegraphics[width=.32\linewidth]{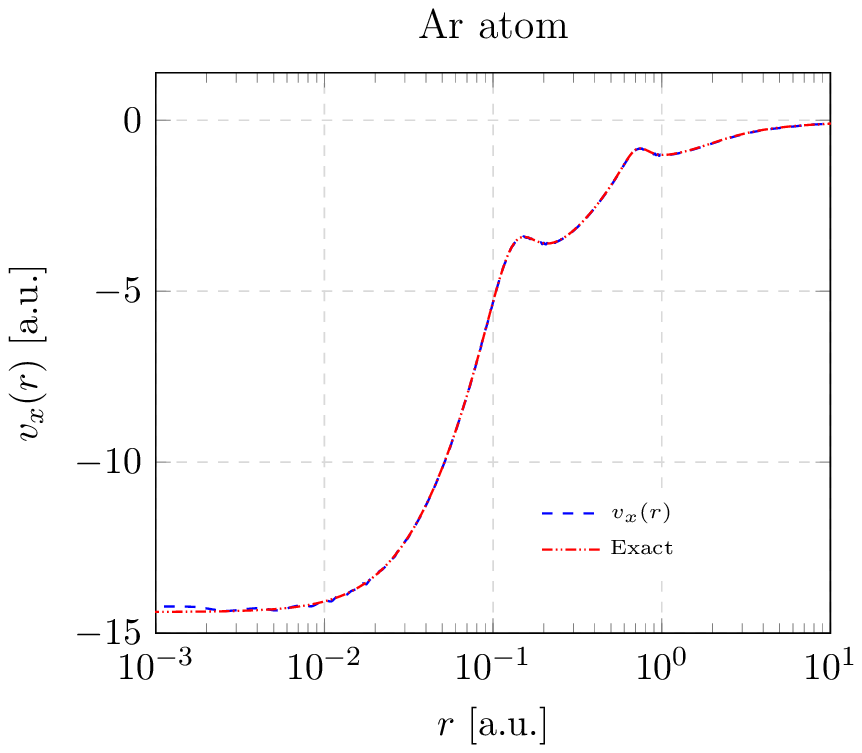}\hfil
\caption{\label{fig3.1} Exchange potentials  $v_{x}(\vect{r})$ for closed shell atom.}
\end{center}
\end{figure*}
\begin{figure*}
\begin{center}
\includegraphics[width=.32\linewidth]{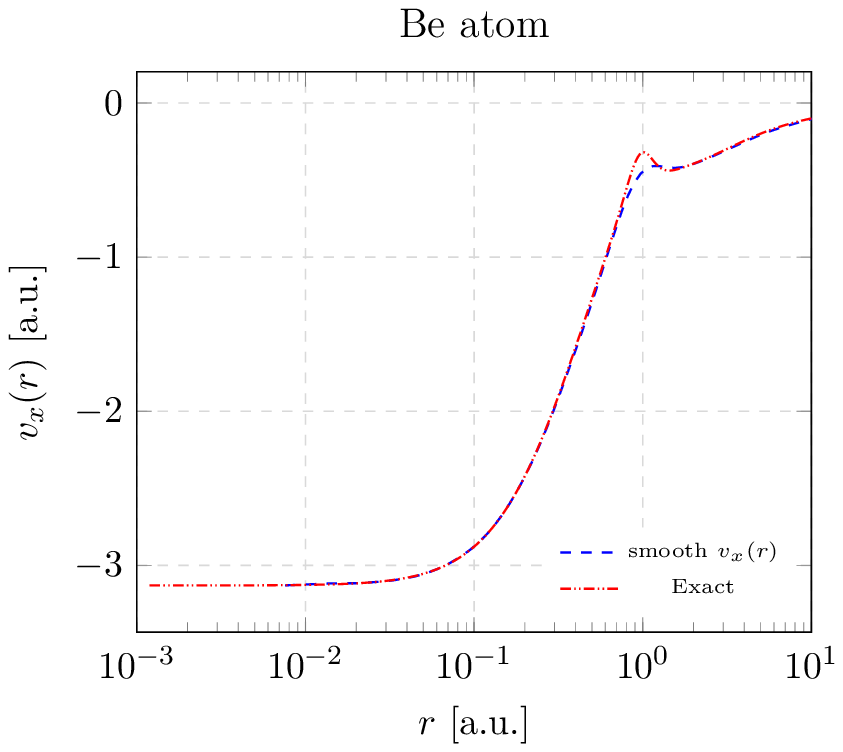}\hfil
\includegraphics[width=.32\linewidth]{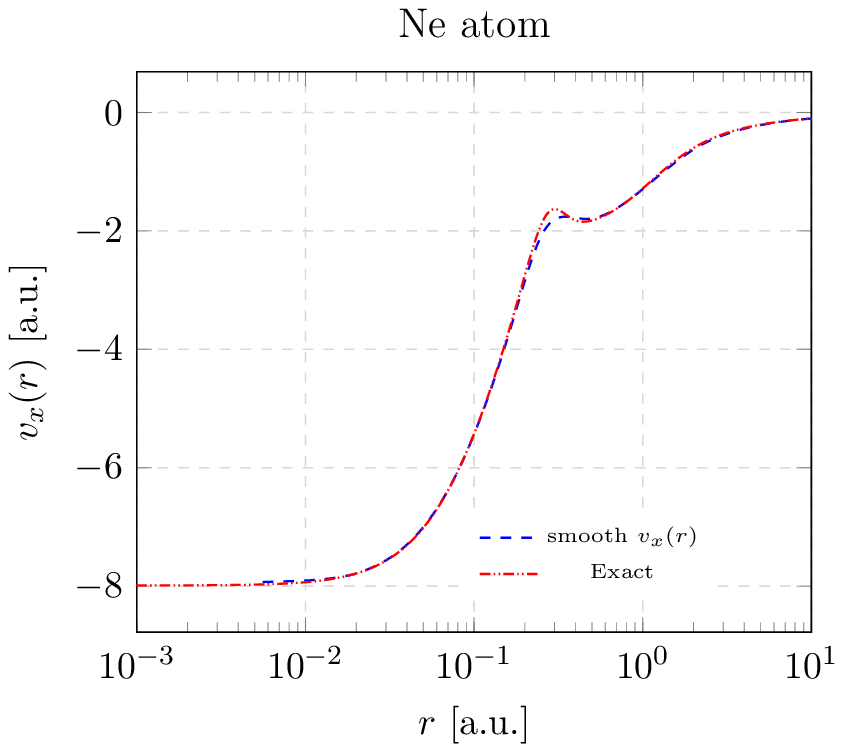}\hfil
\includegraphics[width=.32\linewidth]{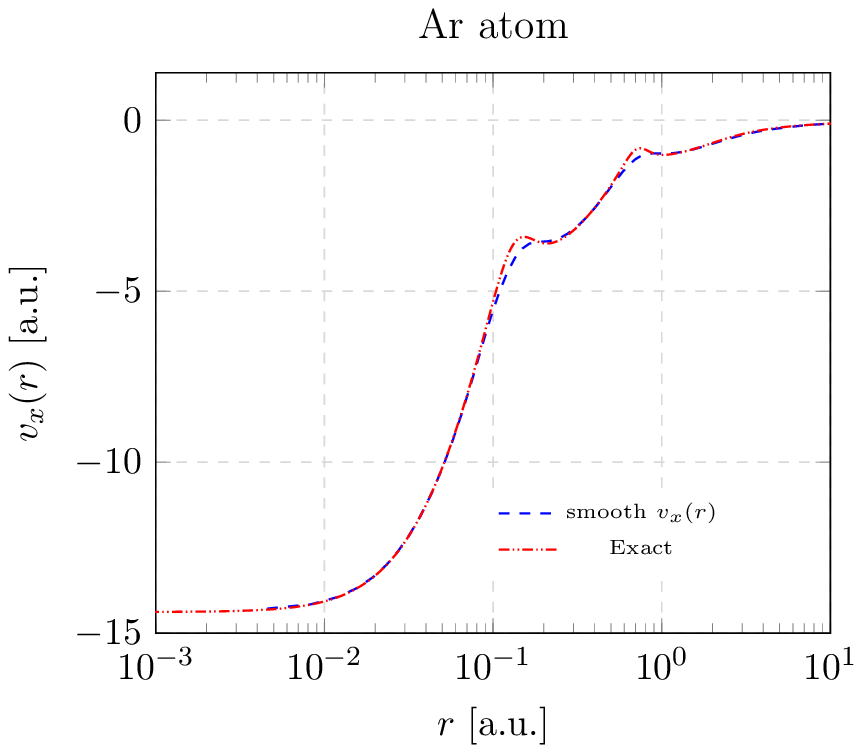}\hfil
\caption{\label{fig3.2} Exchange potentials  $v_{x}(\vect{r})$ after smooth.}
\end{center}
\end{figure*}
\begin{figure*}
\begin{center}
\includegraphics[width=.33\linewidth]{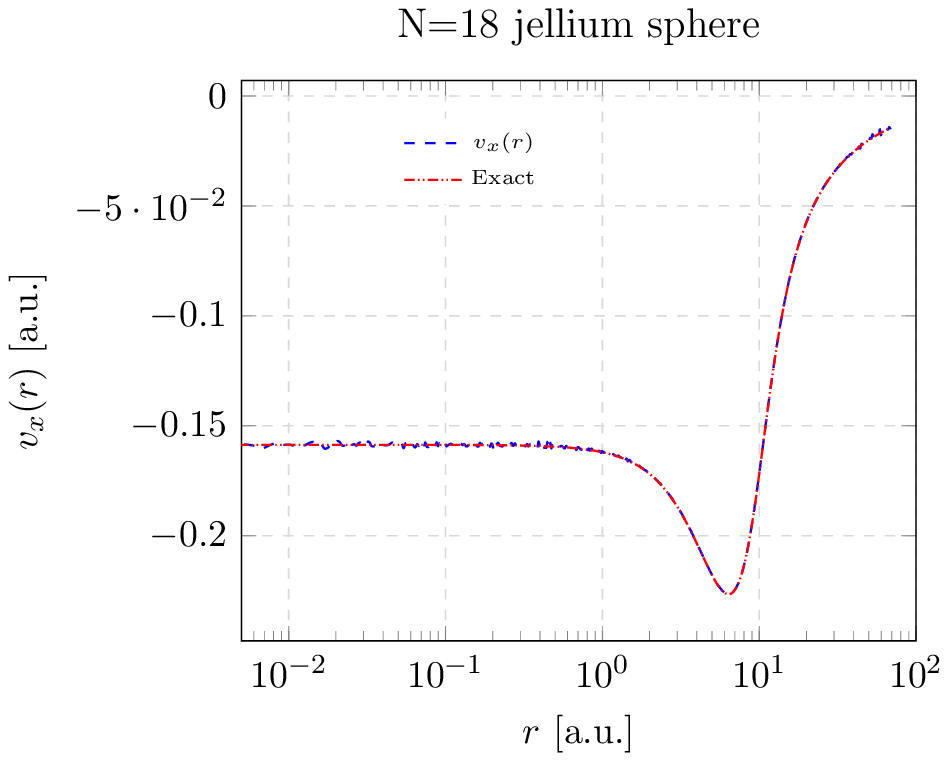}\hfil
\includegraphics[width=.33\linewidth]{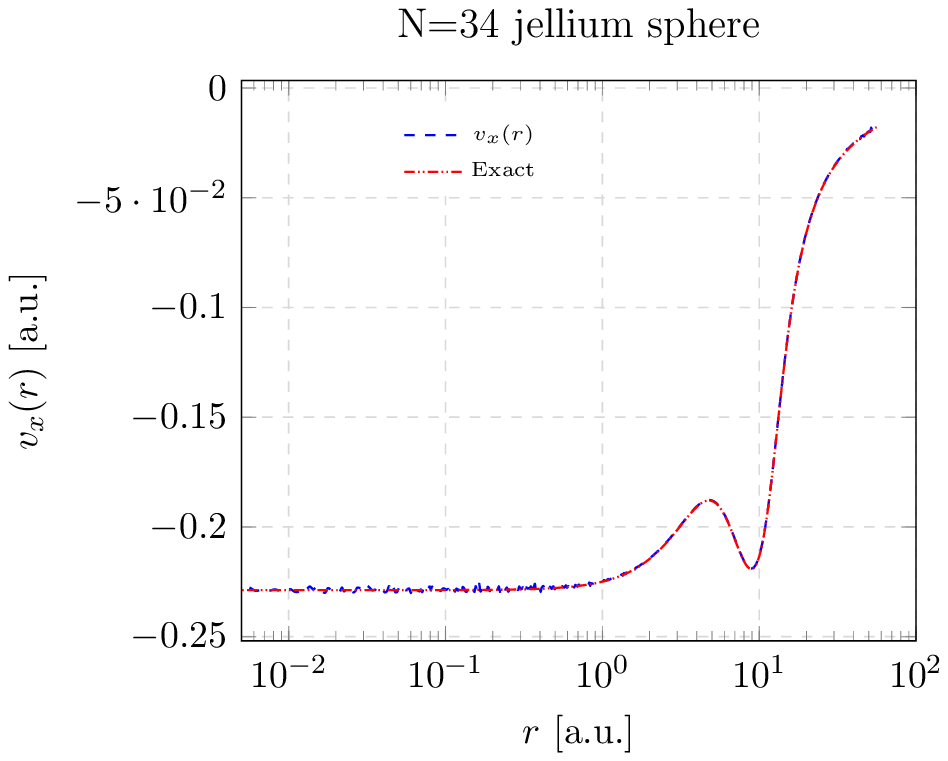}\hfil
\includegraphics[width=.33\linewidth]{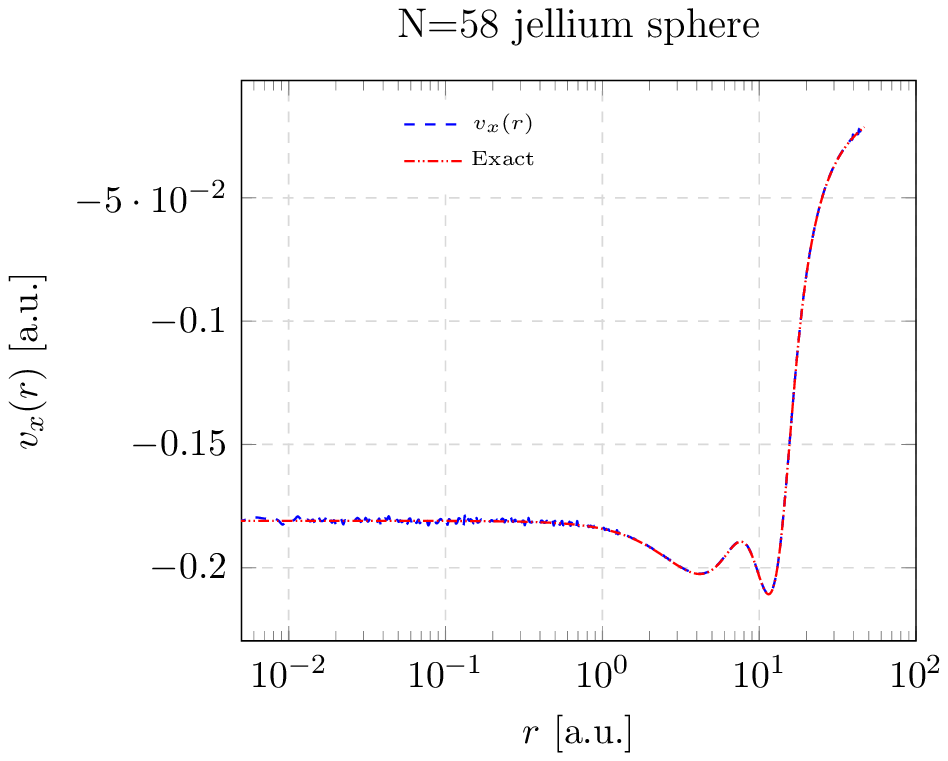}\hfil
\caption{\label{fig3.3} Exchange potentials  $v_{x}(\vect{r})$ for jellium sphere with $N=18, 34 \text{ and } 58$ atoms.}
\end{center}
\end{figure*}
\begin{figure}
\begin{center}
\includegraphics[scale=0.75]{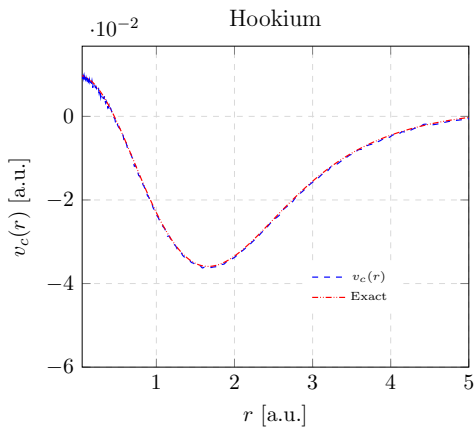} \hfill
\caption{\label{fig3.4} Correlation potential  $v_{c}(\vect{r})$ for Hookium atom.}
\end{center}
\end{figure}
\par In DFT the energy $E[\rho]$ of system of ground state density $\rho(\vect{r})$ is given by
\begin{equation}\label{sec3.2.1}
E[\rho] = F[\rho] +\int v_{ext}(\vect{r})\rho(\vect{r})d\vect{r}
\end{equation}
where $F[\rho]$ is a universal functional of the density and is given by using Levy's constrained search \cite{Levy_1979}
\begin{equation}\label{sec3.2.2}
F[\rho]  = \underset{\Psi \to \rho}{\min} \langle \Psi |T +V_{ee}| \Psi\rangle
\end{equation}
where $T$ is the kinetic energy operator and $V_{ee}$ is the electron-electron interaction energy operator. The search is made over those $|\Psi \rangle$ that are  \text{N-electron} antisymmetric functions  giving the density $\rho(\vect{r})$. The universal functional can also be obtained by finding a potential $v(\vect{r})$ such that
\begin{equation}\label{sec3.2.3}
F[\rho] = \max \left[ E[v] - \int v(\vect{r})\rho(\vect{r}) d\vect{r}\right]
\end{equation}
where $E[v]$ is the energy of \text{N-electrons}  moving in potential $v(\vect{r})$. It is clear that for a given density $\rho(\vect{r})$, the constrained search definition of Eq. (\ref{sec3.2.2}) leads to many-body wavefunction and from it the corresponding potential. Similarly, use of Eq. (\ref{sec3.2.3}) gives the potential corresponding to a given density directly. Thus both of these definitions give a method to invert the density to find the corresponding potential. For example in the work of Teale et al. \cite{Teal2}  they have used Eq. (\ref{sec3.2.3}) to obtain the external potential for a given density for varying strength of electron-electron interaction. The same procedures can be applied to get the Kohn-Sham potential if $F[\rho]$ is treated as the expectation value of the kinetic energy operator and therefore $E[v]$ as the energy of the non-interacting electrons in the potential $v(\vect{r})$. Wu and Yang \cite{WY2} made the first explicit use of Eq. (\ref{sec3.2.3}) to obtain the exchange-correlation potential for a given density. They expanded  the exchange-correlation potential in terms of appropriately chosen Gaussian functions and optimized the coefficients  to maximize  right side of Eq. (\ref{sec3.2.3}).  A general prescription \cite{Kumar_2019} for implementing the approach employing  Eq. (\ref{sec3.2.3}) works as follows. 
\par To find the exchange-correlation potential one starts with an approximate exchange-correlation potential  $v_{xc}^{i}(\vect{r})$, solve Kohn-Sham equation with it, and get a density $ \rho_{i}(\vect{r})$. From the density $ \rho_{i}(\vect{r})$ the exchange-correlation potential  for next iteration, $v_{xc}^{i+1}(\vect{r})$,  is constructed  using the formula
\begin{equation}\label{sec3.2.7}
v_{xc}^{i+1}(\vect{r}) = v_{xc}^{i}(\vect{r}) + \frac{\delta S[\rho]}{\delta \rho(\vect{r})}\Big|_{ \rho_i(\vect{r})} - \frac{\delta S[\rho]}{\delta \rho(\vect{r})}\Big|_{ \rho_0(\vect{r})}
\end{equation}
and solving the corresponding Kohn-Sham equation. This is done  iteratively until a convergence criterion is satisfied. Here $S[\rho]$ is a functional of the dimension of energy and satisfies the condition
\begin{equation}\label{sec3.2.8}
\int \left( \frac{\delta S[\rho]}{\delta \rho(\vect{r})}\Big|_{ \rho_i(\vect{r})} - \frac{\delta S[\rho]}{\delta \rho(\vect{r})}\Big|_{ \rho_0(\vect{r})} \right)\left( \rho_i(\vect{r}) -  \rho_0(\vect{r}) \right) \ge 0.
\end{equation}
As the iterations progress, the exchange-correlation potential becomes close to the true potential and the value of integral in Eq. (\ref{sec3.2.8})  becomes smaller and smaller. It is this procedure that we make use of in employing random numbers to get the exchange-correlation. This is described next.
\par In Eq. (\ref{sec3.2.7})  the correction to the  exchange-correlation potential 
\begin{equation}\label{eq3.3.1}
\Delta v_{xc}^{i+1}(\vect{r})=  \frac{\delta S[\rho]}{\delta \rho(\vect{r})}\Big|_{ \rho_i(\vect{r})} - \frac{\delta S[\rho]}{\delta \rho(\vect{r})}\Big|_{ \rho_0(\vect{r})}
\end{equation}
during the iterations is obtained through the  functional derivative of $S[\rho]$.  Question that we  ask now is if a search can be made for the change $\Delta v_{xc}^{i+1}(\vect{r})$ in the potential directly without being tied down to a functional  $S[\rho]$. The motivation for this being that a fixed functional $S[\rho]$ tends to treat all regions in a system on equal footing, irrespective of the value of the density there. It is, however, found that different functionals forms work better in different regions. Thus a more flexible approach is desirable in this regard. One such method has been to use a hybrid $S[\rho]$ \cite{Kumar_2019}. In the present work we completely abandon the use of a functional $S[\rho]$ of density and propose an up dating scheme that is based on random numbers and therefore fully flexible. It avoids the need to evaluate functional $S[\rho]$ again and again for each iteration.  Furthermore, it provides an advantage over the functional form in those regions where very small  densities make the evaluation of the functional derivative $\frac{\delta S[\rho]}{\delta \rho(\vect{r})}$ rather difficult. The method works as follows.
\par Given as potential $v_{xc}^{i}(\vect{r})$ for the $i^{th}$ iteration, the correction  $\Delta v_{xc}$ added to it to obtain $v_{xc}^{i+1}(\vect{r})$ is constructed using random numbers. For this, at each point of the numerical grid of $(\vect{r})$  we generate random numbers in the range $\{ 0,1\}$, multiply  them by a strength parameter (discussed in the next paragraph ) and choose its sign such that
\begin{equation}\label{eq3.3.2}
 \Delta v_{xc}(\vect{r})(  \rho_i(\vect{r}) -  \rho_0(\vect{r}) ) \ge 0
\end{equation} 
is satisfied. Note that the satisfaction of Eq. (\ref{eq3.3.2}) automatically leads to 
Eq. (\ref{sec3.2.8}) being satisfied. The last step is taken to ensure the convergence of potential towards the correct one by the use of Lieb's definition \cite{Lieb_1983} of $F[\rho]$.  The condition above makes the potential more positive if $(\rho_{i}(\vect{r})-\rho_0(\vect{r}) ) > 0$ and less positive if $(\rho_{i}(\vect{r})-\rho_0(\vect{r}) ) < 0$. The algorithm to generate the exchange-correlation  thus is as follows. We start with an approximate exchange-correlation potential $v_{xc}^0(\vect{r})$ and solve the KS equation with  $v_{ext}(\vect{r})$, exact $v_{H}(\vect{r})$ calculated from the given density $\rho_0(\vect{r})$, and the approximate exchange-correlation potential. In going from $i^{th}$ iteration  to $ (i+1)^{th}$ iteration we keep $v_H(\vect{r})$ fixed as it has been calculated exactly, and update only the exchange-correlation potential. We generate a random profile for $\Delta v_{xc}(\vect{r})$ as follows
\begin{equation}\label{eq3.3.3}
\Delta v_{xc}(\vect{r}) = \lambda f_R(\vect{r}) Sgn[\rho_i,\rho_0](\vect{r}),
\end{equation}
where $f_R(\vect{r})$  takes random values between $0$ and $1$ at each $\vect{r}$. The function $ Sgn[\rho_i,\rho_0](\vect{r})$ is defined as
\begin{equation*}
   Sgn[\rho_i,\rho_0](\vect{r})= 
\begin{cases}
   +1 ,& \text{if } \rho_i(\vect{r}) > \rho_0(\vect{r}) \\
    -1 ,& \text{if } \rho_i(\vect{r}) < \rho_0(\vect{r})
\end{cases}
\end{equation*}
In Eq. ( \ref{eq3.3.3}) $\lambda$ is the strength parameter and is calculated  for each iteration  based on the difference between $\rho_i(\vect{r})$ and $\rho_0(\vect{r})$. For example it could be chosen to be the maximum of  $|\rho_i(\vect{r})-\rho_0(\vect{r})|$. Thus the form of potential upgradation in  Eq. ( \ref{eq3.3.3}) ensures the condition of Eq. (\ref{eq3.3.2}) being satisfied. Using $\Delta v_{xc}^{i+1}(\vect{r})$, the exchange-correlation potential for the next iteration is given as
\begin{equation} \label{eq3.3.4}
v_{xc}^{i+1}(\vect{r}) =(1-\epsilon)v_{xc}^{i}(\vect{r}) +\epsilon(v_{xc}^{i}(\vect{r})+\Delta v_{xc}(\vect{r}) ). 
\end{equation}
where $\epsilon  \quad ( 0 < \epsilon <1)$ is the mixing parameter.  The process is iterated until desired accuracy in density is achieved.
\par We have applied  the method above   to generate the exchange-correlation potential for Hartree-Fock density of atoms  \cite{Bunge_1993} Be, Ne and Ar. Here the external potential is proportional to $- \frac{1}{r}$ where $r$ is the distance from the nucleus. To test applicability of the algorithm for different external potentials  we have also applied it to the Hookium atom \cite{Laufer_1986} and jellium spheres \cite{Knight_PRL.52.2141,Matthias_RMP.65.677}. In the Hookium atom the external potential is proportional to $r^2$ and for the jellium spheres it is proportional to  $r^2$ inside the sphere and proportional to $- \frac{1}{r}$ for $r$ outside the sphere. The potentials calculated by us are compared with the exact results. In our calculations we have chosen parameter $\epsilon$ in Eq. (\ref{eq3.3.4}) to be order of  $10^{-3}-10^{-4}$ and $\lambda$  to be the  maximum of $|\rho_i^n(\vect{r}) -\rho_0^n(\vect{r})|$ for the $i^{th}$ iteration. Here $n=1$ for atoms $n= 0.01$ for the Hookium atom and jellium spheres. The initial  potential $v_{xc}^0(\vect{r}) $ is taken to be  the Fermi-Amaldi potential $- \frac{v_H(\vect{r})}{N}$
 where $N$ is the total number of electrons and $v_H(\vect{r})$ is the Hartree potential corresponding to the  input  density $\rho_0(\vect{r})$. We have also fixed the exchange-correlation potential to its exact value $ -\frac{1}{r}$  \cite{UvBarth_1985} in the asymptotic region. Calculations has been performed using a modified Herman-Skillman code \cite{HSK}. In all the calculations reported we have run the code until  the integral $\int |\rho(\vect{r})- \rho_0(\vect{r})| d\vect{r}  $ becomes smaller than $1.0 \times {10}^{-5}$ for atoms and smaller than $1.0 \times {10}^{-6}$ for jellium spheres and Hookium atom.

\par In Fig (\ref{fig3.1}), we display the exchange potential  for  atoms mentioned above and compare it with the exact exchange potential of these atoms obtained through the optimized potential method  \cite{PhysRev.90.317,Talman_1978,Engel_1993}.   It is evident from the figure that output exchange potential is very close to the corresponding exact results although there are some fluctuations due to the use of random numbers. When these fluctuations are made smooth, the resulting potential becomes essentially exact  as shown in Fig (\ref{fig3.2}). We note, however, that smoothening softens the bump in the intershell region slightly.  Next in Fig. (\ref{fig3.3}), we have plotted the exchange  potential for jellium spheres \cite{Knight_PRL.52.2141,Matthias_RMP.65.677} having $N = 18, 34 \text{ and } 58$ atoms. Here the density of jellium sphere  is obtained by solving the Kohn-Sham equation with the exchange potential taken to be the Harbola-Sahni (HS) potential \cite{MKH_89}. The potentials calculated from the present method are on the top of  the corresponding exact HS potential. Finally in Fig. (\ref{fig3.4}), we have plotted the correlation potential of Hookium atom \cite{Laufer_1986} along with the exact correlation potential. Again the calculated potential for the Hookium atom matches with the exact result.  
\begin{table}
	\caption{\label{tab3.1}   Shown here are the highest occupied  Kohn-Sham  eigenenergy  $\epsilon_{max}$ obtained in this work, the chemical potential $\mu$ which is exact for density employed and    $|\epsilon_{max} - \mu|$. }
	\begin{ruledtabular}
	\begin{tabular}{lcccc}
System  &     &    $\epsilon_{max}$ & $\mu $& $ |\epsilon_{max} - \mu|$   \\ \hline                         
&Be &            -0.311491 & -0.309269&   0.002222  \\
Atom &Ne &              -0.850389&  -0.850410&   0.000021  \\
&Ar &             -0.600938&  -0.591016 &  0.009922 \\
&Hookium &      1.249727&  1.25&  0.000273 \\ \\
&N=18 &           -0.142558&  -0.142692 &  0.000134  \\
jellium spheres&N=34 &           -0.134226 &  -0.134449&  0.000223 \\
&N =58 &           -0.128641&  -0.128775&  0.000134 \\ \\
\end{tabular}
\end{ruledtabular}
\end{table}
\par A hallmark of the accuracy of of an exchange-correlation potential is the satisfaction of  the ionization potential theorem \cite{PPLB,LPS_1984}. We have tested the results of $\epsilon_{max}$ obtained in our calculations against the corresponding exact results. These are shown in Table (\ref{tab3.1}). This is clear that the two are quite close. It is important since we  fix  behaviour of the potential to be $-\frac{1}{r}$ quite far from  the origin; it is the point where the density becomes order of $10^{-6} -10^{-5}$ for atoms and Hookium and less than $10^{-15}$ for  clusters. Thus the potentials of all system calculated by us using the random numbers satisfy the ionization theorem to high degree of accuracy.
\par To conclude, in the present work we have proposed an inversion method to get the exchange-correlation potential for a given density by updating the exchange-correlation potential by employing random numbers. This method circumvents the need to calculate  a functional during the update and thus avoids any difficulties faced in low density region in the calculation of the functional. The method has been applied to different spherical systems and the calculated exchange-correlation potentials are found to be close to exact results.
\begin{acknowledgments}
We are grateful to  Prof. Dr. Eberhard Engel for providing optimized effective potential data of atoms.
\end{acknowledgments}
\bibliography{akmybib}

\begin{thebibliography}{65}%
\makeatletter
\providecommand \@ifxundefined [1]{%
 \@ifx{#1\undefined}
}%
\providecommand \@ifnum [1]{%
 \ifnum #1\expandafter \@firstoftwo
 \else \expandafter \@secondoftwo
 \fi
}%
\providecommand \@ifx [1]{%
 \ifx #1\expandafter \@firstoftwo
 \else \expandafter \@secondoftwo
 \fi
}%
\providecommand \natexlab [1]{#1}%
\providecommand \enquote  [1]{``#1''}%
\providecommand \bibnamefont  [1]{#1}%
\providecommand \bibfnamefont [1]{#1}%
\providecommand \citenamefont [1]{#1}%
\providecommand \href@noop [0]{\@secondoftwo}%
\providecommand \href [0]{\begingroup \@sanitize@url \@href}%
\providecommand \@href[1]{\@@startlink{#1}\@@href}%
\providecommand \@@href[1]{\endgroup#1\@@endlink}%
\providecommand \@sanitize@url [0]{\catcode `\\12\catcode `\$12\catcode
  `\&12\catcode `\#12\catcode `\^12\catcode `\_12\catcode `\%12\relax}%
\providecommand \@@startlink[1]{}%
\providecommand \@@endlink[0]{}%
\providecommand \url  [0]{\begingroup\@sanitize@url \@url }%
\providecommand \@url [1]{\endgroup\@href {#1}{\urlprefix }}%
\providecommand \urlprefix  [0]{URL }%
\providecommand \Eprint [0]{\href }%
\providecommand \doibase [0]{http://dx.doi.org/}%
\providecommand \selectlanguage [0]{\@gobble}%
\providecommand \bibinfo  [0]{\@secondoftwo}%
\providecommand \bibfield  [0]{\@secondoftwo}%
\providecommand \translation [1]{[#1]}%
\providecommand \BibitemOpen [0]{}%
\providecommand \bibitemStop [0]{}%
\providecommand \bibitemNoStop [0]{.\EOS\space}%
\providecommand \EOS [0]{\spacefactor3000\relax}%
\providecommand \BibitemShut  [1]{\csname bibitem#1\endcsname}%
\let\auto@bib@innerbib\@empty
\bibitem [{\citenamefont {Hohenberg}\ and\ \citenamefont {Kohn}(1964)}]{Hohn}%
  \BibitemOpen
  \bibfield  {author} {\bibinfo {author} {\bibfnamefont {P.}~\bibnamefont
  {Hohenberg}}\ and\ \bibinfo {author} {\bibfnamefont {W.}~\bibnamefont
  {Kohn}},\ }\href {https://link.aps.org/doi/10.1103/PhysRev.136.B864}
  {\bibfield  {journal} {\bibinfo  {journal} {Phys. Rev.}\ }\textbf {\bibinfo
  {volume} {136}},\ \bibinfo {pages} {B864} (\bibinfo {year}
  {1964})}\BibitemShut {NoStop}%
\bibitem [{\citenamefont {Kohn}\ and\ \citenamefont {Sham}(1965)}]{Kohn_1965}%
  \BibitemOpen
  \bibfield  {author} {\bibinfo {author} {\bibfnamefont {W.}~\bibnamefont
  {Kohn}}\ and\ \bibinfo {author} {\bibfnamefont {L.~J.}\ \bibnamefont
  {Sham}},\ }\href {https://link.aps.org/doi/10.1103/PhysRev.140.A1133}
  {\bibfield  {journal} {\bibinfo  {journal} {Phys. Rev.}\ }\textbf {\bibinfo
  {volume} {140}},\ \bibinfo {pages} {A1133} (\bibinfo {year}
  {1965})}\BibitemShut {NoStop}%
\bibitem [{\citenamefont {Parr}\ and\ \citenamefont {Yang}(1995)}]{Yang}%
  \BibitemOpen
  \bibfield  {author} {\bibinfo {author} {\bibfnamefont {R.~G.}\ \bibnamefont
  {Parr}}\ and\ \bibinfo {author} {\bibfnamefont {W.}~\bibnamefont {Yang}},\
  }\href@noop {} {\emph {\bibinfo {title} {Density-Functional Theory of Atoms
  and Molecules}}}\ (\bibinfo  {publisher} {Oxford Science Publications},\
  \bibinfo {year} {1995})\BibitemShut {NoStop}%
\bibitem [{\citenamefont {Dreizler}\ and\ \citenamefont {Gross}(1990)}]{Gross}%
  \BibitemOpen
  \bibfield  {author} {\bibinfo {author} {\bibfnamefont {R.~M.}\ \bibnamefont
  {Dreizler}}\ and\ \bibinfo {author} {\bibfnamefont {E.~K.~U.}\ \bibnamefont
  {Gross}},\ }\href@noop {} {\emph {\bibinfo {title} {Density Functional
  Theory}}}\ (\bibinfo  {publisher} {Springer-Verlag Berlin Heidelberg},\
  \bibinfo {year} {1990})\BibitemShut {NoStop}%
\bibitem [{\citenamefont {Engel}\ and\ \citenamefont {Dreizler}(2011)}]{Drei}%
  \BibitemOpen
  \bibfield  {author} {\bibinfo {author} {\bibfnamefont {E.}~\bibnamefont
  {Engel}}\ and\ \bibinfo {author} {\bibfnamefont {R.~M.}\ \bibnamefont
  {Dreizler}},\ }\href@noop {} {\emph {\bibinfo {title} {Density Functional
  Theory}}}\ (\bibinfo  {publisher} {Springer-Verlag Berlin Heidelberg},\
  \bibinfo {year} {2011})\BibitemShut {NoStop}%
\bibitem [{\citenamefont {Pribram-Jones}\ \emph {et~al.}(2015)\citenamefont
  {Pribram-Jones}, \citenamefont {Gross},\ and\ \citenamefont
  {Burke}}]{Burke_1_An_rev_2015}%
  \BibitemOpen
  \bibfield  {author} {\bibinfo {author} {\bibfnamefont {A.}~\bibnamefont
  {Pribram-Jones}}, \bibinfo {author} {\bibfnamefont {D.~A.}\ \bibnamefont
  {Gross}}, \ and\ \bibinfo {author} {\bibfnamefont {K.}~\bibnamefont
  {Burke}},\ }\href {\doibase 10.1146/annurev-physchem-040214-121420}
  {\bibfield  {journal} {\bibinfo  {journal} {Annu. Rev. Phys. Chem.}\ }\textbf
  {\bibinfo {volume} {66}},\ \bibinfo {pages} {283} (\bibinfo {year}
  {2015})}\BibitemShut {NoStop}%
\bibitem [{\citenamefont {Perdew}\ and\ \citenamefont {Yue}(1986)}]{Perdew_86}%
  \BibitemOpen
  \bibfield  {author} {\bibinfo {author} {\bibfnamefont {J.~P.}\ \bibnamefont
  {Perdew}}\ and\ \bibinfo {author} {\bibfnamefont {W.}~\bibnamefont {Yue}},\
  }\href {\doibase 10.1103/PhysRevB.33.8800} {\bibfield  {journal} {\bibinfo
  {journal} {Phys. Rev. B}\ }\textbf {\bibinfo {volume} {33}},\ \bibinfo
  {pages} {8800} (\bibinfo {year} {1986})}\BibitemShut {NoStop}%
\bibitem [{\citenamefont {Becke}(1988)}]{Becke_1988}%
  \BibitemOpen
  \bibfield  {author} {\bibinfo {author} {\bibfnamefont {A.~D.}\ \bibnamefont
  {Becke}},\ }\href {\doibase 10.1103/PhysRevA.38.3098} {\bibfield  {journal}
  {\bibinfo  {journal} {Phys. Rev. A}\ }\textbf {\bibinfo {volume} {38}},\
  \bibinfo {pages} {3098} (\bibinfo {year} {1988})}\BibitemShut {NoStop}%
\bibitem [{\citenamefont {Lee}\ \emph {et~al.}(1988)\citenamefont {Lee},
  \citenamefont {Yang},\ and\ \citenamefont {Parr}}]{Lee_PRB.37.785}%
  \BibitemOpen
  \bibfield  {author} {\bibinfo {author} {\bibfnamefont {C.}~\bibnamefont
  {Lee}}, \bibinfo {author} {\bibfnamefont {W.}~\bibnamefont {Yang}}, \ and\
  \bibinfo {author} {\bibfnamefont {R.~G.}\ \bibnamefont {Parr}},\ }\href
  {https://link.aps.org/doi/10.1103/PhysRevB.37.785} {\bibfield  {journal}
  {\bibinfo  {journal} {Phys. Rev. B}\ }\textbf {\bibinfo {volume} {37}},\
  \bibinfo {pages} {785} (\bibinfo {year} {1988})}\BibitemShut {NoStop}%
\bibitem [{\citenamefont {Perdew}\ \emph {et~al.}(1996)\citenamefont {Perdew},
  \citenamefont {Burke},\ and\ \citenamefont {Ernzerhof}}]{Perdew_PRL.77.3865}%
  \BibitemOpen
  \bibfield  {author} {\bibinfo {author} {\bibfnamefont {J.~P.}\ \bibnamefont
  {Perdew}}, \bibinfo {author} {\bibfnamefont {K.}~\bibnamefont {Burke}}, \
  and\ \bibinfo {author} {\bibfnamefont {M.}~\bibnamefont {Ernzerhof}},\ }\href
  {https://link.aps.org/doi/10.1103/PhysRevLett.77.3865} {\bibfield  {journal}
  {\bibinfo  {journal} {Phys. Rev. Lett.}\ }\textbf {\bibinfo {volume} {77}},\
  \bibinfo {pages} {3865} (\bibinfo {year} {1996})}\BibitemShut {NoStop}%
\bibitem [{\citenamefont {Perdew}\ \emph {et~al.}(1999)\citenamefont {Perdew},
  \citenamefont {Kurth}, \citenamefont {Zupan},\ and\ \citenamefont
  {Blaha}}]{Perdew_PRL.82.2544}%
  \BibitemOpen
  \bibfield  {author} {\bibinfo {author} {\bibfnamefont {J.~P.}\ \bibnamefont
  {Perdew}}, \bibinfo {author} {\bibfnamefont {S.}~\bibnamefont {Kurth}},
  \bibinfo {author} {\bibfnamefont {A.}~\bibnamefont {Zupan}}, \ and\ \bibinfo
  {author} {\bibfnamefont {P.}~\bibnamefont {Blaha}},\ }\href
  {https://link.aps.org/doi/10.1103/PhysRevLett.82.2544} {\bibfield  {journal}
  {\bibinfo  {journal} {Phys. Rev. Lett.}\ }\textbf {\bibinfo {volume} {82}},\
  \bibinfo {pages} {2544} (\bibinfo {year} {1999})}\BibitemShut {NoStop}%
\bibitem [{\citenamefont {Sun}\ \emph {et~al.}(2015{\natexlab{a}})\citenamefont
  {Sun}, \citenamefont {Perdew},\ and\ \citenamefont {Ruzsinszky}}]{Sun_PNAS}%
  \BibitemOpen
  \bibfield  {author} {\bibinfo {author} {\bibfnamefont {J.}~\bibnamefont
  {Sun}}, \bibinfo {author} {\bibfnamefont {J.~P.}\ \bibnamefont {Perdew}}, \
  and\ \bibinfo {author} {\bibfnamefont {A.}~\bibnamefont {Ruzsinszky}},\
  }\href {\doibase 10.1073/pnas.1423145112} {\bibfield  {journal} {\bibinfo
  {journal} {Proc. Natl. Acad. Sci. U.S.A}\ }\textbf {\bibinfo {volume}
  {112}},\ \bibinfo {pages} {685} (\bibinfo {year}
  {2015}{\natexlab{a}})}\BibitemShut {NoStop}%
\bibitem [{\citenamefont {Sun}\ \emph {et~al.}(2015{\natexlab{b}})\citenamefont
  {Sun}, \citenamefont {Ruzsinszky},\ and\ \citenamefont
  {Perdew}}]{Sun_PRL.115.036402}%
  \BibitemOpen
  \bibfield  {author} {\bibinfo {author} {\bibfnamefont {J.}~\bibnamefont
  {Sun}}, \bibinfo {author} {\bibfnamefont {A.}~\bibnamefont {Ruzsinszky}}, \
  and\ \bibinfo {author} {\bibfnamefont {J.~P.}\ \bibnamefont {Perdew}},\
  }\href {https://link.aps.org/doi/10.1103/PhysRevLett.115.036402} {\bibfield
  {journal} {\bibinfo  {journal} {Phys. Rev. Lett.}\ }\textbf {\bibinfo
  {volume} {115}},\ \bibinfo {pages} {036402} (\bibinfo {year}
  {2015}{\natexlab{b}})}\BibitemShut {NoStop}%
\bibitem [{\citenamefont {Sun}\ \emph {et~al.}(2016)\citenamefont {Sun},
  \citenamefont {Remsing}, \citenamefont {Zhang}, \citenamefont {Sun},
  \citenamefont {Ruzsinszky}, \citenamefont {Peng}, \citenamefont {Yang},
  \citenamefont {Paul}, \citenamefont {Waghmare}, \citenamefont {Wu},
  \citenamefont {Klein},\ and\ \citenamefont {Perdew}}]{SCANACC}%
  \BibitemOpen
  \bibfield  {author} {\bibinfo {author} {\bibfnamefont {J.}~\bibnamefont
  {Sun}}, \bibinfo {author} {\bibfnamefont {R.~C.}\ \bibnamefont {Remsing}},
  \bibinfo {author} {\bibfnamefont {Y.}~\bibnamefont {Zhang}}, \bibinfo
  {author} {\bibfnamefont {Z.}~\bibnamefont {Sun}}, \bibinfo {author}
  {\bibfnamefont {A.}~\bibnamefont {Ruzsinszky}}, \bibinfo {author}
  {\bibfnamefont {H.}~\bibnamefont {Peng}}, \bibinfo {author} {\bibfnamefont
  {Z.}~\bibnamefont {Yang}}, \bibinfo {author} {\bibfnamefont {A.}~\bibnamefont
  {Paul}}, \bibinfo {author} {\bibfnamefont {U.}~\bibnamefont {Waghmare}},
  \bibinfo {author} {\bibfnamefont {X.}~\bibnamefont {Wu}}, \bibinfo {author}
  {\bibfnamefont {M.~L.}\ \bibnamefont {Klein}}, \ and\ \bibinfo {author}
  {\bibfnamefont {J.~P.}\ \bibnamefont {Perdew}},\ }\href
  {http://dx.doi.org/10.1038/nchem.2535} {\bibfield  {journal} {\bibinfo
  {journal} {Nat. Chem.}\ }\textbf {\bibinfo {volume} {8}},\ \bibinfo {pages}
  {831} (\bibinfo {year} {2016})}\BibitemShut {NoStop}%
\bibitem [{\citenamefont {Tao}\ \emph {et~al.}(2003)\citenamefont {Tao},
  \citenamefont {Perdew}, \citenamefont {Staroverov},\ and\ \citenamefont
  {Scuseria}}]{Jianmin_PRL.91.146401}%
  \BibitemOpen
  \bibfield  {author} {\bibinfo {author} {\bibfnamefont {J.}~\bibnamefont
  {Tao}}, \bibinfo {author} {\bibfnamefont {J.~P.}\ \bibnamefont {Perdew}},
  \bibinfo {author} {\bibfnamefont {V.~N.}\ \bibnamefont {Staroverov}}, \ and\
  \bibinfo {author} {\bibfnamefont {G.~E.}\ \bibnamefont {Scuseria}},\ }\href
  {https://link.aps.org/doi/10.1103/PhysRevLett.91.146401} {\bibfield
  {journal} {\bibinfo  {journal} {Phys. Rev. Lett.}\ }\textbf {\bibinfo
  {volume} {91}},\ \bibinfo {pages} {146401} (\bibinfo {year}
  {2003})}\BibitemShut {NoStop}%
\bibitem [{\citenamefont {Almbladh}\ and\ \citenamefont {von
  Barth}(1985)}]{UvBarth_1985}%
  \BibitemOpen
  \bibfield  {author} {\bibinfo {author} {\bibfnamefont {C.-O.}\ \bibnamefont
  {Almbladh}}\ and\ \bibinfo {author} {\bibfnamefont {U.}~\bibnamefont {von
  Barth}},\ }\href {\doibase 10.1103/PhysRevB.31.3231} {\bibfield  {journal}
  {\bibinfo  {journal} {Phys. Rev. B}\ }\textbf {\bibinfo {volume} {31}},\
  \bibinfo {pages} {3231} (\bibinfo {year} {1985})}\BibitemShut {NoStop}%
\bibitem [{\citenamefont {Perdew}\ \emph {et~al.}(1982)\citenamefont {Perdew},
  \citenamefont {Parr}, \citenamefont {Levy},\ and\ \citenamefont {Jr}}]{PPLB}%
  \BibitemOpen
  \bibfield  {author} {\bibinfo {author} {\bibfnamefont {J.~P.}\ \bibnamefont
  {Perdew}}, \bibinfo {author} {\bibfnamefont {R.~G.}\ \bibnamefont {Parr}},
  \bibinfo {author} {\bibfnamefont {M.}~\bibnamefont {Levy}}, \ and\ \bibinfo
  {author} {\bibfnamefont {J.~L.~B.}\ \bibnamefont {Jr}},\ }\href
  {http://journals.aps.org/prl/pdf/10.1103/PhysRevLett.49.1691} {\bibfield
  {journal} {\bibinfo  {journal} {Phys. Rev. Lett.}\ }\textbf {\bibinfo
  {volume} {49}},\ \bibinfo {pages} {1691} (\bibinfo {year}
  {1982})}\BibitemShut {NoStop}%
\bibitem [{\citenamefont {Levy}\ \emph {et~al.}(1984)\citenamefont {Levy},
  \citenamefont {Perdew},\ and\ \citenamefont {Sahni}}]{LPS_1984}%
  \BibitemOpen
  \bibfield  {author} {\bibinfo {author} {\bibfnamefont {M.}~\bibnamefont
  {Levy}}, \bibinfo {author} {\bibfnamefont {J.~P.}\ \bibnamefont {Perdew}}, \
  and\ \bibinfo {author} {\bibfnamefont {V.}~\bibnamefont {Sahni}},\ }\href
  {\doibase 10.1103/PhysRevA.30.2745} {\bibfield  {journal} {\bibinfo
  {journal} {Phys. Rev. A}\ }\textbf {\bibinfo {volume} {30}},\ \bibinfo
  {pages} {2745} (\bibinfo {year} {1984})}\BibitemShut {NoStop}%
\bibitem [{\citenamefont {Buijse}\ \emph {et~al.}(1989)\citenamefont {Buijse},
  \citenamefont {Baerends},\ and\ \citenamefont {Snijders}}]{Buijse_1989}%
  \BibitemOpen
  \bibfield  {author} {\bibinfo {author} {\bibfnamefont {M.~A.}\ \bibnamefont
  {Buijse}}, \bibinfo {author} {\bibfnamefont {E.~J.}\ \bibnamefont
  {Baerends}}, \ and\ \bibinfo {author} {\bibfnamefont {J.~G.}\ \bibnamefont
  {Snijders}},\ }\href {\doibase 10.1103/PhysRevA.40.4190} {\bibfield
  {journal} {\bibinfo  {journal} {Phys. Rev. A}\ }\textbf {\bibinfo {volume}
  {40}},\ \bibinfo {pages} {4190} (\bibinfo {year} {1989})}\BibitemShut
  {NoStop}%
\bibitem [{\citenamefont {Gritsenko}\ and\ \citenamefont
  {Baerends}(1996)}]{Gritsenko_1996}%
  \BibitemOpen
  \bibfield  {author} {\bibinfo {author} {\bibfnamefont {O.~V.}\ \bibnamefont
  {Gritsenko}}\ and\ \bibinfo {author} {\bibfnamefont {E.~J.}\ \bibnamefont
  {Baerends}},\ }\href {\doibase 10.1103/PhysRevA.54.1957} {\bibfield
  {journal} {\bibinfo  {journal} {Phys. Rev. A}\ }\textbf {\bibinfo {volume}
  {54}},\ \bibinfo {pages} {1957} (\bibinfo {year} {1996})}\BibitemShut
  {NoStop}%
\bibitem [{\citenamefont {Teale}\ \emph {et~al.}(2009)\citenamefont {Teale},
  \citenamefont {Coriani},\ and\ \citenamefont {Helgaker}}]{Teal2}%
  \BibitemOpen
  \bibfield  {author} {\bibinfo {author} {\bibfnamefont {A.~M.}\ \bibnamefont
  {Teale}}, \bibinfo {author} {\bibfnamefont {S.}~\bibnamefont {Coriani}}, \
  and\ \bibinfo {author} {\bibfnamefont {T.}~\bibnamefont {Helgaker}},\ }\href
  {\doibase 10.1063/1.3082285} {\bibfield  {journal} {\bibinfo  {journal} {J.
  Chem. Phys.}\ }\textbf {\bibinfo {volume} {130}},\ \bibinfo {pages} {104111}
  (\bibinfo {year} {2009})}\BibitemShut {NoStop}%
\bibitem [{\citenamefont {Teale}\ \emph
  {et~al.}(2010{\natexlab{a}})\citenamefont {Teale}, \citenamefont {Coriani},\
  and\ \citenamefont {Helgaker}}]{Teal3}%
  \BibitemOpen
  \bibfield  {author} {\bibinfo {author} {\bibfnamefont {A.~M.}\ \bibnamefont
  {Teale}}, \bibinfo {author} {\bibfnamefont {S.}~\bibnamefont {Coriani}}, \
  and\ \bibinfo {author} {\bibfnamefont {T.}~\bibnamefont {Helgaker}},\ }\href
  {\doibase 10.1063/1.3380834} {\bibfield  {journal} {\bibinfo  {journal} {J.
  Chem. Phys.}\ }\textbf {\bibinfo {volume} {132}},\ \bibinfo {pages} {164115}
  (\bibinfo {year} {2010}{\natexlab{a}})}\BibitemShut {NoStop}%
\bibitem [{\citenamefont {Teale}\ \emph
  {et~al.}(2010{\natexlab{b}})\citenamefont {Teale}, \citenamefont {Coriani},\
  and\ \citenamefont {Helgaker}}]{Teal4}%
  \BibitemOpen
  \bibfield  {author} {\bibinfo {author} {\bibfnamefont {A.~M.}\ \bibnamefont
  {Teale}}, \bibinfo {author} {\bibfnamefont {S.}~\bibnamefont {Coriani}}, \
  and\ \bibinfo {author} {\bibfnamefont {T.}~\bibnamefont {Helgaker}},\ }\href
  {\doibase 10.1063/1.3488100} {\bibfield  {journal} {\bibinfo  {journal} {J.
  Chem. Phys.}\ }\textbf {\bibinfo {volume} {133}},\ \bibinfo {pages} {164112}
  (\bibinfo {year} {2010}{\natexlab{b}})}\BibitemShut {NoStop}%
\bibitem [{\citenamefont {Makmal}\ \emph {et~al.}(2011)\citenamefont {Makmal},
  \citenamefont {K\"ummel},\ and\ \citenamefont {Kronik}}]{Makmal_2011}%
  \BibitemOpen
  \bibfield  {author} {\bibinfo {author} {\bibfnamefont {A.}~\bibnamefont
  {Makmal}}, \bibinfo {author} {\bibfnamefont {S.}~\bibnamefont {K\"ummel}}, \
  and\ \bibinfo {author} {\bibfnamefont {L.}~\bibnamefont {Kronik}},\ }\href
  {\doibase 10.1103/PhysRevA.83.062512} {\bibfield  {journal} {\bibinfo
  {journal} {Phys. Rev. A}\ }\textbf {\bibinfo {volume} {83}},\ \bibinfo
  {pages} {062512} (\bibinfo {year} {2011})}\BibitemShut {NoStop}%
\bibitem [{\citenamefont {Stoudenmire}\ \emph {et~al.}(2012)\citenamefont
  {Stoudenmire}, \citenamefont {Wagner}, \citenamefont {White},\ and\
  \citenamefont {Burke}}]{Wagner_2012}%
  \BibitemOpen
  \bibfield  {author} {\bibinfo {author} {\bibfnamefont {E.~M.}\ \bibnamefont
  {Stoudenmire}}, \bibinfo {author} {\bibfnamefont {L.~O.}\ \bibnamefont
  {Wagner}}, \bibinfo {author} {\bibfnamefont {S.~R.}\ \bibnamefont {White}}, \
  and\ \bibinfo {author} {\bibfnamefont {K.}~\bibnamefont {Burke}},\ }\href
  {\doibase 10.1103/PhysRevLett.109.056402} {\bibfield  {journal} {\bibinfo
  {journal} {Phys. Rev. Lett.}\ }\textbf {\bibinfo {volume} {109}},\ \bibinfo
  {pages} {056402} (\bibinfo {year} {2012})}\BibitemShut {NoStop}%
\bibitem [{\citenamefont {Gould}\ and\ \citenamefont
  {Toulouse}(2014)}]{2014_Gould}%
  \BibitemOpen
  \bibfield  {author} {\bibinfo {author} {\bibfnamefont {T.}~\bibnamefont
  {Gould}}\ and\ \bibinfo {author} {\bibfnamefont {J.}~\bibnamefont
  {Toulouse}},\ }\href {\doibase 10.1103/PhysRevA.90.050502} {\bibfield
  {journal} {\bibinfo  {journal} {Phys. Rev. A}\ }\textbf {\bibinfo {volume}
  {90}},\ \bibinfo {pages} {050502} (\bibinfo {year} {2014})}\BibitemShut
  {NoStop}%
\bibitem [{\citenamefont {Kohut}\ \emph {et~al.}(2016)\citenamefont {Kohut},
  \citenamefont {Polgar},\ and\ \citenamefont {Staroverov}}]{Kohut_2016}%
  \BibitemOpen
  \bibfield  {author} {\bibinfo {author} {\bibfnamefont {S.~V.}\ \bibnamefont
  {Kohut}}, \bibinfo {author} {\bibfnamefont {A.~M.}\ \bibnamefont {Polgar}}, \
  and\ \bibinfo {author} {\bibfnamefont {V.~N.}\ \bibnamefont {Staroverov}},\
  }\href {\doibase 10.1039/C6CP00878J} {\bibfield  {journal} {\bibinfo
  {journal} {Phys. Chem. Chem. Phys.}\ }\textbf {\bibinfo {volume} {18}},\
  \bibinfo {pages} {20938} (\bibinfo {year} {2016})}\BibitemShut {NoStop}%
\bibitem [{\citenamefont {Ben\'{\i}tez}\ and\ \citenamefont
  {Proetto}(2016)}]{Proetto_2016}%
  \BibitemOpen
  \bibfield  {author} {\bibinfo {author} {\bibfnamefont {A.}~\bibnamefont
  {Ben\'{\i}tez}}\ and\ \bibinfo {author} {\bibfnamefont {C.~R.}\ \bibnamefont
  {Proetto}},\ }\href {\doibase 10.1103/PhysRevA.94.052506} {\bibfield
  {journal} {\bibinfo  {journal} {Phys. Rev. A}\ }\textbf {\bibinfo {volume}
  {94}},\ \bibinfo {pages} {052506} (\bibinfo {year} {2016})}\BibitemShut
  {NoStop}%
\bibitem [{\citenamefont {Hodgson}\ \emph {et~al.}(2016)\citenamefont
  {Hodgson}, \citenamefont {Ramsden},\ and\ \citenamefont
  {Godby}}]{Godby_PRA_2016}%
  \BibitemOpen
  \bibfield  {author} {\bibinfo {author} {\bibfnamefont {M.~J.~P.}\
  \bibnamefont {Hodgson}}, \bibinfo {author} {\bibfnamefont {J.~D.}\
  \bibnamefont {Ramsden}}, \ and\ \bibinfo {author} {\bibfnamefont {R.~W.}\
  \bibnamefont {Godby}},\ }\href {\doibase 10.1103/PhysRevB.93.155146}
  {\bibfield  {journal} {\bibinfo  {journal} {Phys. Rev. B}\ }\textbf {\bibinfo
  {volume} {93}},\ \bibinfo {pages} {155146} (\bibinfo {year}
  {2016})}\BibitemShut {NoStop}%
\bibitem [{\citenamefont {Singh}\ and\ \citenamefont
  {Harbola}(2017)}]{Rabeet_2017}%
  \BibitemOpen
  \bibfield  {author} {\bibinfo {author} {\bibfnamefont {R.}~\bibnamefont
  {Singh}}\ and\ \bibinfo {author} {\bibfnamefont {M.~K.}\ \bibnamefont
  {Harbola}},\ }\href {\doibase 10.1063/1.4995698} {\bibfield  {journal}
  {\bibinfo  {journal} {J. Chem. Phys.}\ }\textbf {\bibinfo {volume} {147}},\
  \bibinfo {pages} {144105} (\bibinfo {year} {2017})}\BibitemShut {NoStop}%
\bibitem [{\citenamefont {Ospadov}\ \emph {et~al.}(2018)\citenamefont
  {Ospadov}, \citenamefont {Tao}, \citenamefont {Staroverov},\ and\
  \citenamefont {Perdew}}]{Staroverov_PNAS_18}%
  \BibitemOpen
  \bibfield  {author} {\bibinfo {author} {\bibfnamefont {E.}~\bibnamefont
  {Ospadov}}, \bibinfo {author} {\bibfnamefont {J.}~\bibnamefont {Tao}},
  \bibinfo {author} {\bibfnamefont {V.~N.}\ \bibnamefont {Staroverov}}, \ and\
  \bibinfo {author} {\bibfnamefont {J.~P.}\ \bibnamefont {Perdew}},\ }\href
  {\doibase 10.1073/pnas.1814300115} {\bibfield  {journal} {\bibinfo  {journal}
  {Proc. Natl. Acad. Sci. U.S.A}\ }\textbf {\bibinfo {volume} {115}},\ \bibinfo
  {pages} {E11578} (\bibinfo {year} {2018})}\BibitemShut {NoStop}%
\bibitem [{\citenamefont {Kaiser}\ and\ \citenamefont
  {K\"ummel}(2018)}]{kummel_18}%
  \BibitemOpen
  \bibfield  {author} {\bibinfo {author} {\bibfnamefont {A.}~\bibnamefont
  {Kaiser}}\ and\ \bibinfo {author} {\bibfnamefont {S.}~\bibnamefont
  {K\"ummel}},\ }\href {\doibase 10.1103/PhysRevA.98.052505} {\bibfield
  {journal} {\bibinfo  {journal} {Phys. Rev. A}\ }\textbf {\bibinfo {volume}
  {98}},\ \bibinfo {pages} {052505} (\bibinfo {year} {2018})}\BibitemShut
  {NoStop}%
\bibitem [{\citenamefont {Gould}\ \emph {et~al.}(2019)\citenamefont {Gould},
  \citenamefont {Pittalis}, \citenamefont {Toulouse}, \citenamefont
  {Kraisler},\ and\ \citenamefont {Kronik}}]{2019_Gould}%
  \BibitemOpen
  \bibfield  {author} {\bibinfo {author} {\bibfnamefont {T.}~\bibnamefont
  {Gould}}, \bibinfo {author} {\bibfnamefont {S.}~\bibnamefont {Pittalis}},
  \bibinfo {author} {\bibfnamefont {J.}~\bibnamefont {Toulouse}}, \bibinfo
  {author} {\bibfnamefont {E.}~\bibnamefont {Kraisler}}, \ and\ \bibinfo
  {author} {\bibfnamefont {L.}~\bibnamefont {Kronik}},\ }\href {\doibase
  10.1039/C9CP03633D} {\bibfield  {journal} {\bibinfo  {journal} {Phys. Chem.
  Chem. Phys.}\ }\textbf {\bibinfo {volume} {21}},\ \bibinfo {pages} {19805}
  (\bibinfo {year} {2019})}\BibitemShut {NoStop}%
\bibitem [{\citenamefont {Gritsenko}\ \emph {et~al.}(1998)\citenamefont
  {Gritsenko}, \citenamefont {Schipper},\ and\ \citenamefont
  {Baerends}}]{Gritsenko_1998}%
  \BibitemOpen
  \bibfield  {author} {\bibinfo {author} {\bibfnamefont {O.~V.}\ \bibnamefont
  {Gritsenko}}, \bibinfo {author} {\bibfnamefont {P.~R.~T.}\ \bibnamefont
  {Schipper}}, \ and\ \bibinfo {author} {\bibfnamefont {E.~J.}\ \bibnamefont
  {Baerends}},\ }\href {\doibase 10.1103/PhysRevA.57.3450} {\bibfield
  {journal} {\bibinfo  {journal} {Phys. Rev. A}\ }\textbf {\bibinfo {volume}
  {57}},\ \bibinfo {pages} {3450} (\bibinfo {year} {1998})}\BibitemShut
  {NoStop}%
\bibitem [{\citenamefont {Schipper}\ \emph {et~al.}(1998)\citenamefont
  {Schipper}, \citenamefont {Gritsenko},\ and\ \citenamefont
  {Baerends}}]{Schipper_1998}%
  \BibitemOpen
  \bibfield  {author} {\bibinfo {author} {\bibfnamefont {P.~R.~T.}\
  \bibnamefont {Schipper}}, \bibinfo {author} {\bibfnamefont {O.~V.}\
  \bibnamefont {Gritsenko}}, \ and\ \bibinfo {author} {\bibfnamefont {E.~J.}\
  \bibnamefont {Baerends}},\ }\href {\doibase 10.1103/PhysRevA.57.1729}
  {\bibfield  {journal} {\bibinfo  {journal} {Phys. Rev. A}\ }\textbf {\bibinfo
  {volume} {57}},\ \bibinfo {pages} {1729} (\bibinfo {year}
  {1998})}\BibitemShut {NoStop}%
\bibitem [{\citenamefont {Ryabinkin}\ \emph {et~al.}(2013)\citenamefont
  {Ryabinkin}, \citenamefont {Kananenka},\ and\ \citenamefont
  {Staroverov}}]{Viktor_2013}%
  \BibitemOpen
  \bibfield  {author} {\bibinfo {author} {\bibfnamefont {I.~G.}\ \bibnamefont
  {Ryabinkin}}, \bibinfo {author} {\bibfnamefont {A.~A.}\ \bibnamefont
  {Kananenka}}, \ and\ \bibinfo {author} {\bibfnamefont {V.~N.}\ \bibnamefont
  {Staroverov}},\ }\href {\doibase 10.1103/PhysRevLett.111.013001} {\bibfield
  {journal} {\bibinfo  {journal} {Phys. Rev. Lett.}\ }\textbf {\bibinfo
  {volume} {111}},\ \bibinfo {pages} {013001} (\bibinfo {year}
  {2013})}\BibitemShut {NoStop}%
\bibitem [{\citenamefont {Ryabinkin}\ \emph {et~al.}(2015)\citenamefont
  {Ryabinkin}, \citenamefont {Kohut},\ and\ \citenamefont
  {Staroverov}}]{Viktor_2015}%
  \BibitemOpen
  \bibfield  {author} {\bibinfo {author} {\bibfnamefont {I.~G.}\ \bibnamefont
  {Ryabinkin}}, \bibinfo {author} {\bibfnamefont {S.~V.}\ \bibnamefont
  {Kohut}}, \ and\ \bibinfo {author} {\bibfnamefont {V.~N.}\ \bibnamefont
  {Staroverov}},\ }\href {\doibase 10.1103/PhysRevLett.115.083001} {\bibfield
  {journal} {\bibinfo  {journal} {Phys. Rev. Lett.}\ }\textbf {\bibinfo
  {volume} {115}},\ \bibinfo {pages} {083001} (\bibinfo {year}
  {2015})}\BibitemShut {NoStop}%
\bibitem [{\citenamefont {Werden}\ and\ \citenamefont
  {Davidson}(1984)}]{Werden}%
  \BibitemOpen
  \bibfield  {author} {\bibinfo {author} {\bibfnamefont {S.~H.}\ \bibnamefont
  {Werden}}\ and\ \bibinfo {author} {\bibfnamefont {E.~R.}\ \bibnamefont
  {Davidson}},\ }in\ \href {\doibase https://doi.org/10.1007/978-1-4899-2142-0}
  {\emph {\bibinfo {booktitle} {Local Density Approximations in Quantum
  Chemistry and Solid State Physics}}},\ \bibinfo {editor} {edited by\ \bibinfo
  {editor} {\bibfnamefont {J.~P.}\ \bibnamefont {Dahl}}\ and\ \bibinfo {editor}
  {\bibfnamefont {J.}~\bibnamefont {Avery}}}\ (\bibinfo  {publisher} {Springer,
  Boston, MA},\ \bibinfo {year} {1984})\ Chap.\ \bibinfo {chapter} {On the
  Calculation of Potentials from Densities}\BibitemShut {NoStop}%
\bibitem [{\citenamefont {Aryasetiawan}\ and\ \citenamefont
  {Stott}(1988)}]{Stott_1988}%
  \BibitemOpen
  \bibfield  {author} {\bibinfo {author} {\bibfnamefont {F.}~\bibnamefont
  {Aryasetiawan}}\ and\ \bibinfo {author} {\bibfnamefont {M.~J.}\ \bibnamefont
  {Stott}},\ }\href {\doibase 10.1103/PhysRevB.38.2974} {\bibfield  {journal}
  {\bibinfo  {journal} {Phys. Rev. B}\ }\textbf {\bibinfo {volume} {38}},\
  \bibinfo {pages} {2974} (\bibinfo {year} {1988})}\BibitemShut {NoStop}%
\bibitem [{\citenamefont {G\"orling}(1992)}]{Gorling_1992}%
  \BibitemOpen
  \bibfield  {author} {\bibinfo {author} {\bibfnamefont {A.}~\bibnamefont
  {G\"orling}},\ }\href {\doibase 10.1103/PhysRevA.46.3753} {\bibfield
  {journal} {\bibinfo  {journal} {Phys. Rev. A}\ }\textbf {\bibinfo {volume}
  {46}},\ \bibinfo {pages} {3753} (\bibinfo {year} {1992})}\BibitemShut
  {NoStop}%
\bibitem [{\citenamefont {Zhao}\ and\ \citenamefont {Parr}(1992)}]{Zhao_1992}%
  \BibitemOpen
  \bibfield  {author} {\bibinfo {author} {\bibfnamefont {Q.}~\bibnamefont
  {Zhao}}\ and\ \bibinfo {author} {\bibfnamefont {R.~G.}\ \bibnamefont
  {Parr}},\ }\href {\doibase 10.1103/PhysRevA.46.2337} {\bibfield  {journal}
  {\bibinfo  {journal} {Phys. Rev. A}\ }\textbf {\bibinfo {volume} {46}},\
  \bibinfo {pages} {2337} (\bibinfo {year} {1992})}\BibitemShut {NoStop}%
\bibitem [{\citenamefont {Wang}\ and\ \citenamefont {Parr}(1993)}]{Wang_1993}%
  \BibitemOpen
  \bibfield  {author} {\bibinfo {author} {\bibfnamefont {Y.}~\bibnamefont
  {Wang}}\ and\ \bibinfo {author} {\bibfnamefont {R.~G.}\ \bibnamefont
  {Parr}},\ }\href {\doibase 10.1103/PhysRevA.47.R1591} {\bibfield  {journal}
  {\bibinfo  {journal} {Phys. Rev. A}\ }\textbf {\bibinfo {volume} {47}},\
  \bibinfo {pages} {R1591} (\bibinfo {year} {1993})}\BibitemShut {NoStop}%
\bibitem [{\citenamefont {Zhao}\ and\ \citenamefont {Parr}(1993)}]{Zhao_1993}%
  \BibitemOpen
  \bibfield  {author} {\bibinfo {author} {\bibfnamefont {Q.}~\bibnamefont
  {Zhao}}\ and\ \bibinfo {author} {\bibfnamefont {R.~G.}\ \bibnamefont
  {Parr}},\ }\href {\doibase 10.1063/1.465093} {\bibfield  {journal} {\bibinfo
  {journal} {J. Chem. Phys.}\ }\textbf {\bibinfo {volume} {98}},\ \bibinfo
  {pages} {543} (\bibinfo {year} {1993})}\BibitemShut {NoStop}%
\bibitem [{\citenamefont {Zhao}\ \emph {et~al.}(1994)\citenamefont {Zhao},
  \citenamefont {Morrison},\ and\ \citenamefont {Parr}}]{Zhao_1994}%
  \BibitemOpen
  \bibfield  {author} {\bibinfo {author} {\bibfnamefont {Q.}~\bibnamefont
  {Zhao}}, \bibinfo {author} {\bibfnamefont {R.~C.}\ \bibnamefont {Morrison}},
  \ and\ \bibinfo {author} {\bibfnamefont {R.~G.}\ \bibnamefont {Parr}},\
  }\href {\doibase 10.1103/PhysRevA.50.2138} {\bibfield  {journal} {\bibinfo
  {journal} {Phys. Rev. A}\ }\textbf {\bibinfo {volume} {50}},\ \bibinfo
  {pages} {2138} (\bibinfo {year} {1994})}\BibitemShut {NoStop}%
\bibitem [{\citenamefont {van Leeuwen}\ and\ \citenamefont
  {Baerends}(1994)}]{Vlb_1994}%
  \BibitemOpen
  \bibfield  {author} {\bibinfo {author} {\bibfnamefont {R.}~\bibnamefont {van
  Leeuwen}}\ and\ \bibinfo {author} {\bibfnamefont {E.~J.}\ \bibnamefont
  {Baerends}},\ }\href {\doibase 10.1103/PhysRevA.49.2421} {\bibfield
  {journal} {\bibinfo  {journal} {Phys. Rev. A}\ }\textbf {\bibinfo {volume}
  {49}},\ \bibinfo {pages} {2421} (\bibinfo {year} {1994})}\BibitemShut
  {NoStop}%
\bibitem [{\citenamefont {Schipper}\ \emph {et~al.}(1997)\citenamefont
  {Schipper}, \citenamefont {Gritsenko},\ and\ \citenamefont
  {Baerends}}]{Schipper1997}%
  \BibitemOpen
  \bibfield  {author} {\bibinfo {author} {\bibfnamefont {P.~R.~T.}\
  \bibnamefont {Schipper}}, \bibinfo {author} {\bibfnamefont {O.~V.}\
  \bibnamefont {Gritsenko}}, \ and\ \bibinfo {author} {\bibfnamefont {E.~J.}\
  \bibnamefont {Baerends}},\ }\href {\doibase 10.1007/s002140050273} {\bibfield
   {journal} {\bibinfo  {journal} {Theor. Chem. Acc.}\ }\textbf {\bibinfo
  {volume} {98}},\ \bibinfo {pages} {16} (\bibinfo {year} {1997})}\BibitemShut
  {NoStop}%
\bibitem [{\citenamefont {Wu}\ and\ \citenamefont {Yang}(2003)}]{WY2}%
  \BibitemOpen
  \bibfield  {author} {\bibinfo {author} {\bibfnamefont {Q.}~\bibnamefont
  {Wu}}\ and\ \bibinfo {author} {\bibfnamefont {W.}~\bibnamefont {Yang}},\
  }\href {\doibase 10.1063/1.1535422} {\bibfield  {journal} {\bibinfo
  {journal} {J. Chem. Phys.}\ }\textbf {\bibinfo {volume} {118}},\ \bibinfo
  {pages} {2498} (\bibinfo {year} {2003})}\BibitemShut {NoStop}%
\bibitem [{\citenamefont {Peirs}\ \emph {et~al.}(2003)\citenamefont {Peirs},
  \citenamefont {Van~Neck},\ and\ \citenamefont {Waroquier}}]{Peirs_2003}%
  \BibitemOpen
  \bibfield  {author} {\bibinfo {author} {\bibfnamefont {K.}~\bibnamefont
  {Peirs}}, \bibinfo {author} {\bibfnamefont {D.}~\bibnamefont {Van~Neck}}, \
  and\ \bibinfo {author} {\bibfnamefont {M.}~\bibnamefont {Waroquier}},\ }\href
  {\doibase 10.1103/PhysRevA.67.012505} {\bibfield  {journal} {\bibinfo
  {journal} {Phys. Rev. A}\ }\textbf {\bibinfo {volume} {67}},\ \bibinfo
  {pages} {012505} (\bibinfo {year} {2003})}\BibitemShut {NoStop}%
\bibitem [{\citenamefont {Kadantsev}\ and\ \citenamefont
  {Stott}(2004)}]{Stott_2004}%
  \BibitemOpen
  \bibfield  {author} {\bibinfo {author} {\bibfnamefont {E.~S.}\ \bibnamefont
  {Kadantsev}}\ and\ \bibinfo {author} {\bibfnamefont {M.~J.}\ \bibnamefont
  {Stott}},\ }\href {\doibase 10.1103/PhysRevA.69.012502} {\bibfield  {journal}
  {\bibinfo  {journal} {Phys. Rev. A}\ }\textbf {\bibinfo {volume} {69}},\
  \bibinfo {pages} {012502} (\bibinfo {year} {2004})}\BibitemShut {NoStop}%
\bibitem [{\citenamefont {Wagner}\ \emph {et~al.}(2014)\citenamefont {Wagner},
  \citenamefont {Baker}, \citenamefont {Stoudenmire}, \citenamefont {Burke},\
  and\ \citenamefont {White}}]{Wagner_2014}%
  \BibitemOpen
  \bibfield  {author} {\bibinfo {author} {\bibfnamefont {L.~O.}\ \bibnamefont
  {Wagner}}, \bibinfo {author} {\bibfnamefont {T.~E.}\ \bibnamefont {Baker}},
  \bibinfo {author} {\bibfnamefont {E.~M.}\ \bibnamefont {Stoudenmire}},
  \bibinfo {author} {\bibfnamefont {K.}~\bibnamefont {Burke}}, \ and\ \bibinfo
  {author} {\bibfnamefont {S.~R.}\ \bibnamefont {White}},\ }\href {\doibase
  10.1103/PhysRevB.90.045109} {\bibfield  {journal} {\bibinfo  {journal} {Phys.
  Rev. B}\ }\textbf {\bibinfo {volume} {90}},\ \bibinfo {pages} {045109}
  (\bibinfo {year} {2014})}\BibitemShut {NoStop}%
\bibitem [{\citenamefont {Hollins}\ \emph {et~al.}(2017)\citenamefont
  {Hollins}, \citenamefont {Clark}, \citenamefont {Refson},\ and\ \citenamefont
  {Gidopoulos}}]{Hollins_2017}%
  \BibitemOpen
  \bibfield  {author} {\bibinfo {author} {\bibfnamefont {T.~W.}\ \bibnamefont
  {Hollins}}, \bibinfo {author} {\bibfnamefont {S.~J.}\ \bibnamefont {Clark}},
  \bibinfo {author} {\bibfnamefont {K.}~\bibnamefont {Refson}}, \ and\ \bibinfo
  {author} {\bibfnamefont {N.~I.}\ \bibnamefont {Gidopoulos}},\ }\href
  {http://stacks.iop.org/0953-8984/29/i=4/a=04LT01} {\bibfield  {journal}
  {\bibinfo  {journal} {J. Phys. Condens. Matter}\ }\textbf {\bibinfo {volume}
  {29}},\ \bibinfo {pages} {04LT01} (\bibinfo {year} {2017})}\BibitemShut
  {NoStop}%
\bibitem [{\citenamefont {Jensen}\ and\ \citenamefont
  {Wasserman}(2017)}]{Wasserman_2017}%
  \BibitemOpen
  \bibfield  {author} {\bibinfo {author} {\bibfnamefont {D.~S.}\ \bibnamefont
  {Jensen}}\ and\ \bibinfo {author} {\bibfnamefont {A.}~\bibnamefont
  {Wasserman}},\ }\href {http://dx.doi.org/10.1002/qua.25425} {\bibfield
  {journal} {\bibinfo  {journal} {Int. J. Quantum Chem.}\ } (\bibinfo {year}
  {2017})}\BibitemShut {NoStop}%
\bibitem [{\citenamefont {Finzel}\ \emph {et~al.}(2018)\citenamefont {Finzel},
  \citenamefont {Ayers},\ and\ \citenamefont {Bultinck}}]{Finzel2018}%
  \BibitemOpen
  \bibfield  {author} {\bibinfo {author} {\bibfnamefont {K.}~\bibnamefont
  {Finzel}}, \bibinfo {author} {\bibfnamefont {P.~W.}\ \bibnamefont {Ayers}}, \
  and\ \bibinfo {author} {\bibfnamefont {P.}~\bibnamefont {Bultinck}},\ }\href
  {\doibase 10.1007/s00214-018-2209-0} {\bibfield  {journal} {\bibinfo
  {journal} {Theor. Chem. Acc.}\ }\textbf {\bibinfo {volume} {137}},\ \bibinfo
  {pages} {30} (\bibinfo {year} {2018})}\BibitemShut {NoStop}%
\bibitem [{\citenamefont {Kumar}\ \emph {et~al.}(2019)\citenamefont {Kumar},
  \citenamefont {Singh},\ and\ \citenamefont {Harbola}}]{Kumar_2019}%
  \BibitemOpen
  \bibfield  {author} {\bibinfo {author} {\bibfnamefont {A.}~\bibnamefont
  {Kumar}}, \bibinfo {author} {\bibfnamefont {R.}~\bibnamefont {Singh}}, \ and\
  \bibinfo {author} {\bibfnamefont {M.~K.}\ \bibnamefont {Harbola}},\ }\href
  {\doibase 10.1088/1361-6455/ab04e8} {\bibfield  {journal} {\bibinfo
  {journal} {J. Phys. B: At., Mol. Opt. Phys.}\ }\textbf {\bibinfo {volume}
  {52}},\ \bibinfo {pages} {075007} (\bibinfo {year} {2019})}\BibitemShut
  {NoStop}%
\bibitem [{\citenamefont {Lieb}(1983)}]{Lieb_1983}%
  \BibitemOpen
  \bibfield  {author} {\bibinfo {author} {\bibfnamefont {E.~H.}\ \bibnamefont
  {Lieb}},\ }\href {\doibase 10.1002/qua.560240302} {\bibfield  {journal}
  {\bibinfo  {journal} {Int. J. Quantum Chem.}\ }\textbf {\bibinfo {volume}
  {24}},\ \bibinfo {pages} {243} (\bibinfo {year} {1983})}\BibitemShut
  {NoStop}%
\bibitem [{\citenamefont {Levy}(1979)}]{Levy_1979}%
  \BibitemOpen
  \bibfield  {author} {\bibinfo {author} {\bibfnamefont {M.}~\bibnamefont
  {Levy}},\ }\href {http://www.pnas.org/content/76/12/6062.abstract} {\bibfield
   {journal} {\bibinfo  {journal} {Proc. Natl. Acad. Sci. U.S.A}\ }\textbf
  {\bibinfo {volume} {76}},\ \bibinfo {pages} {6062} (\bibinfo {year}
  {1979})}\BibitemShut {NoStop}%
\bibitem [{\citenamefont {Bunge}\ \emph {et~al.}(1993)\citenamefont {Bunge},
  \citenamefont {Barrientos},\ and\ \citenamefont {Bunge}}]{Bunge_1993}%
  \BibitemOpen
  \bibfield  {author} {\bibinfo {author} {\bibfnamefont {C.}~\bibnamefont
  {Bunge}}, \bibinfo {author} {\bibfnamefont {J.}~\bibnamefont {Barrientos}}, \
  and\ \bibinfo {author} {\bibfnamefont {A.}~\bibnamefont {Bunge}},\ }\href
  {\doibase http://dx.doi.org/10.1006/adnd.1993.1003} {\bibfield  {journal}
  {\bibinfo  {journal} {Atomic Data and Nuclear Data Tables}\ }\textbf
  {\bibinfo {volume} {53}},\ \bibinfo {pages} {113 } (\bibinfo {year}
  {1993})}\BibitemShut {NoStop}%
\bibitem [{\citenamefont {Laufer}\ and\ \citenamefont
  {Krieger}(1986)}]{Laufer_1986}%
  \BibitemOpen
  \bibfield  {author} {\bibinfo {author} {\bibfnamefont {P.~M.}\ \bibnamefont
  {Laufer}}\ and\ \bibinfo {author} {\bibfnamefont {J.~B.}\ \bibnamefont
  {Krieger}},\ }\href {\doibase 10.1103/PhysRevA.33.1480} {\bibfield  {journal}
  {\bibinfo  {journal} {Phys. Rev. A}\ }\textbf {\bibinfo {volume} {33}},\
  \bibinfo {pages} {1480} (\bibinfo {year} {1986})}\BibitemShut {NoStop}%
\bibitem [{\citenamefont {Knight}\ \emph {et~al.}(1984)\citenamefont {Knight},
  \citenamefont {Clemenger}, \citenamefont {de~Heer}, \citenamefont {Saunders},
  \citenamefont {Chou},\ and\ \citenamefont {Cohen}}]{Knight_PRL.52.2141}%
  \BibitemOpen
  \bibfield  {author} {\bibinfo {author} {\bibfnamefont {W.~D.}\ \bibnamefont
  {Knight}}, \bibinfo {author} {\bibfnamefont {K.}~\bibnamefont {Clemenger}},
  \bibinfo {author} {\bibfnamefont {W.~A.}\ \bibnamefont {de~Heer}}, \bibinfo
  {author} {\bibfnamefont {W.~A.}\ \bibnamefont {Saunders}}, \bibinfo {author}
  {\bibfnamefont {M.~Y.}\ \bibnamefont {Chou}}, \ and\ \bibinfo {author}
  {\bibfnamefont {M.~L.}\ \bibnamefont {Cohen}},\ }\href {\doibase
  10.1103/PhysRevLett.52.2141} {\bibfield  {journal} {\bibinfo  {journal}
  {Phys. Rev. Lett.}\ }\textbf {\bibinfo {volume} {52}},\ \bibinfo {pages}
  {2141} (\bibinfo {year} {1984})}\BibitemShut {NoStop}%
\bibitem [{\citenamefont {Brack}(1993)}]{Matthias_RMP.65.677}%
  \BibitemOpen
  \bibfield  {author} {\bibinfo {author} {\bibfnamefont {M.}~\bibnamefont
  {Brack}},\ }\href {\doibase 10.1103/RevModPhys.65.677} {\bibfield  {journal}
  {\bibinfo  {journal} {Rev. Mod. Phys.}\ }\textbf {\bibinfo {volume} {65}},\
  \bibinfo {pages} {677} (\bibinfo {year} {1993})}\BibitemShut {NoStop}%
\bibitem [{\citenamefont {Herman}\ and\ \citenamefont {Skillman}(1963)}]{HSK}%
  \BibitemOpen
  \bibfield  {author} {\bibinfo {author} {\bibfnamefont {F.}~\bibnamefont
  {Herman}}\ and\ \bibinfo {author} {\bibfnamefont {S.}~\bibnamefont
  {Skillman}},\ }\href@noop {} {\emph {\bibinfo {title} {Atomic structure
  calculations}}}\ (\bibinfo  {publisher} {Prentice-Hall Publication},\
  \bibinfo {year} {1963})\BibitemShut {NoStop}%
\bibitem [{\citenamefont {Sharp}\ and\ \citenamefont
  {Horton}(1953)}]{PhysRev.90.317}%
  \BibitemOpen
  \bibfield  {author} {\bibinfo {author} {\bibfnamefont {R.~T.}\ \bibnamefont
  {Sharp}}\ and\ \bibinfo {author} {\bibfnamefont {G.~K.}\ \bibnamefont
  {Horton}},\ }\href {\doibase 10.1103/PhysRev.90.317} {\bibfield  {journal}
  {\bibinfo  {journal} {Phys. Rev.}\ }\textbf {\bibinfo {volume} {90}},\
  \bibinfo {pages} {317} (\bibinfo {year} {1953})}\BibitemShut {NoStop}%
\bibitem [{\citenamefont {Aashamar}\ \emph {et~al.}(1978)\citenamefont
  {Aashamar}, \citenamefont {Luke},\ and\ \citenamefont
  {Talman}}]{Talman_1978}%
  \BibitemOpen
  \bibfield  {author} {\bibinfo {author} {\bibfnamefont {K.}~\bibnamefont
  {Aashamar}}, \bibinfo {author} {\bibfnamefont {T.}~\bibnamefont {Luke}}, \
  and\ \bibinfo {author} {\bibfnamefont {J.}~\bibnamefont {Talman}},\ }\href
  {\doibase http://dx.doi.org/10.1016/0092-640X(78)90019-0} {\bibfield
  {journal} {\bibinfo  {journal} {Atomic Data and Nuclear Data Tables}\
  }\textbf {\bibinfo {volume} {22}},\ \bibinfo {pages} {443 } (\bibinfo {year}
  {1978})}\BibitemShut {NoStop}%
\bibitem [{\citenamefont {Engel}\ and\ \citenamefont
  {Vosko}(1993)}]{Engel_1993}%
  \BibitemOpen
  \bibfield  {author} {\bibinfo {author} {\bibfnamefont {E.}~\bibnamefont
  {Engel}}\ and\ \bibinfo {author} {\bibfnamefont {S.~H.}\ \bibnamefont
  {Vosko}},\ }\href {\doibase 10.1103/PhysRevA.47.2800} {\bibfield  {journal}
  {\bibinfo  {journal} {Phys. Rev. A}\ }\textbf {\bibinfo {volume} {47}},\
  \bibinfo {pages} {2800} (\bibinfo {year} {1993})}\BibitemShut {NoStop}%
\bibitem [{\citenamefont {Harbola}\ and\ \citenamefont {Sahni}(1989)}]{MKH_89}%
  \BibitemOpen
  \bibfield  {author} {\bibinfo {author} {\bibfnamefont {M.~K.}\ \bibnamefont
  {Harbola}}\ and\ \bibinfo {author} {\bibfnamefont {V.}~\bibnamefont
  {Sahni}},\ }\href {\doibase 10.1103/PhysRevLett.62.489} {\bibfield  {journal}
  {\bibinfo  {journal} {Phys. Rev. Lett.}\ }\textbf {\bibinfo {volume} {62}},\
  \bibinfo {pages} {489} (\bibinfo {year} {1989})}\BibitemShut {NoStop}%
\end{thebibliography}%
\end{document}